\documentclass{ws-ijmpd}
\usepackage{amssymb,amsmath}

\newcommand{\be}{\begin{eqnarray}}
\newcommand{\ee}{\end{eqnarray}}

\begin{document}

\title{GRAVITOMAGNETISM IN SUPERCONDUCTORS AND COMPACT STARS}

\author{COSIMO BAMBI}
\address{Department of Physics and Astronomy, 
Wayne State University \\ 
Detroit, MI 48201, USA }

\maketitle

\begin{abstract}
There are three experimentally observed effects in rotating 
superconductors that are so far unexplained. Some authors
have tried to interpret such a phenomena as possible
new gravitational properties of coherent quantum systems:
in particular, they suggest that the gravitomagnetic 
field of that kind of matter may be many orders of
magnitude stronger than the one expected in the standard
theory. Here I show that this interpretation would
be in conflict with the common belief that neutron stars 
have neutrons in superfluid state and protons in 
superconductive one.
\end{abstract}

\section{Introduction}

Gravitomagnetic phenomena in superconductors were
considered for the first time by DeWitt in Ref.~\refcite{dewitt}.
In the 90's, Li and Torr suggested the fascinating possibility 
that superconductors were able to produce anomalous 
strong gravitomagnetic fields\cite{th} and at that time
experiments seemed to support their idea\cite{exp}. Today, 
there are three apparently unexplainable laboratory measurements
on rotating superconductors\cite{london}$^,$\cite{tajmar1} which
have been addressed as a possible indication of new 
anomalous gravitational properties of coherent quantum 
systems (see~\refcite{prop1}, \refcite{prop2} and reference therein).
The latter may be also connected with the nature of the 
so-called dark energy\cite{beck}. 
Indeed, the observed effects would require gravitomagnetic 
fields of many order of magnitude larger than the 
one predicted by general relativity. The issue is that 
general relativity is well tested for classical objects 
like stars, planets and satellites, while these 
mysterious phenomena are seen only when the matter is 
in superconductive or superfluid state, 
below some critical temperature $T_c$.

In this paper I show that such a possibility is quite 
probably in conflict with the physics of very compact objects.
According to the commonly accepted picture of neutron 
stars, the high density and low temperature of the matter
inside these objects cause the neutrons to form $^3P_2$ 
Cooper pairs and to condensate to a superfluid state
and the small fraction of protons to form $^1S_0$ Cooper
pairs and to condensate to a superconductive state\cite{sedrakian}. 
Estimating the surface temperature of some young neutron 
stars, X-ray satellites support this picture\cite{x-ray}.
In addition to this, there is the possibility that inside
neutron stars there are superfluid hyperons\cite{super-h}
and/or superfluid and superconductive quark matter\cite{super-q}.
If this accepted picture were true and superconductors and 
superfluids produced anomalous stronger gravitomagnetic fields,
at the levels suggested by laboratory experiments, gravitomagnetic
effects like the Lense-Thirring one should affect the
motion of neutron star in binary systems as well. This is 
not seen and thus one of the above hypothesis is 
probably wrong. Even if the fraction of superfluid and
superconductive matter inside neutron stars is quite
model dependent, and hence it is difficult to find reliable
constraints on new physics, the gravitomagnetic field suggested
by laboratory experiments are 20 -- 30 orders of magnitude
stronger than the one expected by general relativity, so 
gravitomagnetic phenomena in binary systems should be 
observable even in the most pessimistic scenarios, because 
the standard theory predicts gravitomagnetic effects 
only about 5 orders of magnitude weaker than the main 
gravitoelectric ones and in any case the fraction of superfluid
and superconductive matter is not less than some percent. 
Moreover, laboratory tests\cite{mass} show also that 
superfluids and superconductors have usual gravitoelectric 
properties, so the orbital motion of neutron stars would be 
affected only by anomalous large gravitomagnetic contributions.

The content of the work is the following. In section~\ref{s-gem},
I review the gravitoelectromagnetism framework and, in
section~\ref{s-lab}, the unexplained observed effects
in rotating superconductors. In section~\ref{s-astro},
I discuss the possibility that superfluids or superconductors 
in compact stars produce gravitomagnetic fields as strong 
as one could expect from laboratory experiments and I 
show that this would be inconsistent with observations. In 
section~\ref{s-conclusion}, I report the conclusions of this 
work.

\section{Gravitoelectromagnetism \label{s-gem}}

In the weak field and slow motion approximation, 
Einstein gravitational field equations become formally 
equivalent to the ones of the electromagnetic field: 
we can define a gravitoelectric field ${\bf E}_g$
and a gravitomagnetic field ${\bf B}_g$ which satisfy
Maxwell-like equations
\be
\nabla \cdot {\bf E}_g &=& 4 \pi G_N \, \rho \, , \\
\nabla \cdot \left(\frac{1}{2} {\bf B}_g\right) &=& 0 \, , \\
\nabla \wedge {\bf E}_g &=& - \frac{1}{c} 
\frac{\partial}{\partial t} 
\left(\frac{1}{2} {\bf B}_g\right) \, , \\
\nabla \wedge \left(\frac{1}{2} {\bf B}_g\right) &=& 
\frac{1}{c} \frac{\partial}{\partial t} {\bf E}_g
+ \frac{4 \pi G_N}{c} {\bf j} \, ,
\ee
where $\rho$ is the matter density and ${\bf j}$ is the 
matter current. The factors $1/2$ in front of the
gravitomagnetic field ${\bf B}_g$ arise from the spin-2
nature of the gravitational field. For an introduction
on the gravitoelectromagnetic framework, see e.g.
Ref.~\refcite{gem}.

Reminding the electromagnetic theory, it is easy to
see that the gravitomagnetic field of a rotating body
is proportional to its proper angular momentum ${\bf J}$
and induces the spin precession for any orbiting spinning
test-particle (the so-called Lense-Thirring effect). The 
angular velocity of such a precession is
\be
{\bf \Omega}_{LT} = - \frac{G_N}{c^2} 
\frac{{\bf J} - 3 \, \hat{\bf r} \, 
(\hat{\bf r} \cdot {\bf J})}{r^3} \, ,
\ee
where ${\bf r}$ is the position vector of the test-particle
with respect to the rotating body, $r = |{\bf r}|$ and 
$\hat{\bf r} = {\bf r}/r$.
Just like in the electromagnetic case, one can also consider
the whole orbit of the test-particle as a giant gyroscope
affected by the precession of the longitude of the ascending 
node $\Omega$ and of the argument of the pericenter $\omega$.
This causes a precession of the longitude of the pericenter
$\bar{\omega} = \Omega + \omega$ which is usually much smaller 
than the well known main gravitoelectric effect (and indeed at 
present we cannot observe it in the orbit of Solar System 
planets). The angular velocity of the pericenter precession is
\be\label{o-precession}
\dot{\bar{\omega}} = \frac{2 G_N}{c^2}
\frac{{\bf J} - 3 \, \hat{\bf L} \,
(\hat{\bf L} \cdot {\bf J})}{a^3 (1 - e^2)^{3/2}} \, .
\ee
Here $\hat{\bf L}$ is the unit vector orbital angular momentum 
of the test-particle and $a$ and $e$ are respectively the
semimajor axis and the eccentricity of the orbit of the test-particle.
Evidences of the Earth's gravitomagnetic field have been
reported in Ref.~\refcite{ciufolini} from an analysis of the
laser ranged data of the satellites LAGEOS and LAGEOS~II
and represent one of the main targets of the Gravity Probe B
mission\cite{g-p-b}.

\section{Laboratory Experiments \label{s-lab}}

It is well known that the gravitational force is much
weaker than the electromagnetic and nuclear ones. This
is certainly true for classical ordinary matter, where
general relativity predictions are in agreement with
observational evidences, but it may not be so for coherent
quantum systems. Indeed, some observed effects in 
laboratory experiments may suggest that this kind of 
matter produces much stronger gravitomagnetic fields.

One of the unexplained result is the measurement of
Cooper pair mass in rotating niobium superconductive
rings\cite{london}
\be
\Delta m = m_{Exp} - m_{Th} =
94.147249(21) \, {\rm eV} \, , 
\ee
where $m_{Exp}$ is the measured mass and $m_{Th}$ is
the theoretically predicted one. Such a measurement
could be explained including the gravitomagnetic term
in the canonical momentum of the Cooper pair
\be
\Pi = m {\bf v} + \frac{e}{c} \, {\bf A} 
+ \frac{m}{c} \, {\bf A}_g \, ,
\ee
where ${\bf A}$ and ${\bf A}_g$ are respectively the
electromagnetic and the gravitoelectromagnetic vector
potential (that is ${\bf B}_g = \nabla \wedge {\bf A}_g$).
However, the mass excess can be explained only if the 
gravitomagnetic field ${\bf B}_g$ is 30 orders of
magnitude larger than the one predicted by the standard
theory: indeed experimentally one would deduce\cite{prop1}
\be
\frac{|{\bf B}_g|}{\omega_{ring}} \Big|_{Exp} 
\sim c \, \frac{\Delta m}{m_{Th}} \sim 10^6 \, {\rm cm/s} \, ,
\ee
while the standard theory would predict
\be
\frac{|{\bf B}_g|}{\omega_{ring}} \Big|_{Th} 
\sim \frac{G_N m_{ring}}{c \, R} \sim 10^{-24} \, {\rm cm/s} \, ,
\ee
as one can obtain straightforward from dimensional arguments.
Here $m_{ring} = 2 \, {\rm \mu g}$ is the mass of the
rings in the experiment and $R = 5 \, {\rm cm}$ its radius.

Other two unexplained outcomes in experiments
involving rotating superconductive rings have been
performed at the Austrian Research Centers\cite{tajmar1}. 
One of them measures the azimuthal acceleration $g$ in 
the central hole of different rotating superconductive 
rings and finds an anomalous acceleration directly 
proportional to the ring angular acceleration 
$\dot{\omega}_{ring}$, with the coupling between $g$
and $\dot{\omega}_{ring}$ depending on the superconductor
but typically at the level\cite{tajmar1}
\be
\frac{g}{\dot{\omega}_{ring}} \Big|_{Exp} 
\sim - 10^{-4} \, {\rm cm} \, .
\ee
Even this result can be interpreted as a gravitoelectromagnetic
phenomenon, because a time varying gravitomagnetic field
must induce a gravitoelectric field (Faraday-like induction
law)\cite{prop2}. However, the predicted effect in the
standard theory is completely negligible: a rough
estimate suggests
\be
\frac{g}{\dot{\omega}_{ring}} \Big|_{Th} 
\sim - \frac{G_N m_{ring}}{c^2} \sim - 10^{-26} \, {\rm cm} \, ,
\ee
where $m_{ring} = 350 \, {\rm g}$ is the mass of the
rings in the experiment. In order to account for the
observed effect, the gravitomagnetic field of the superconductor
should be about 22 orders of magnitude larger than its
expected value.

The second experiment at the Austrian Research Centers 
measures the phase difference between
two beams of coherent electromagnetic radiation
with the same frequency $\nu_0$ and propagating in
opposite directions along a closed optical fiber. The 
experiment finds a coupling constant between the phase 
difference $\Delta \varphi$ and the constant angular 
velocity $\omega_{ring}$ of a rotating niobium superconductive 
ring in the neighborhood. The phase difference induced by 
a gravitomagnetic field ${\bf B}_g$ would be
\be
\Delta \varphi = \frac{4 \, \nu_0}{c^3} 
\, {\bf S} \cdot {\bf B}_g \, ,
\ee 
where ${\bf S}$ is the area vector of the optical fiber whose
direction is orthogonal to the fiber plane. If we
interpret such a phase shift as a gravitomagnetic phenomenon,
we find the relation\cite{tajmar1}
\be
\frac{|{\bf B}_g|}{\omega_{ring}} \Big|_{Exp} 
\sim 10^2 \, {\rm cm/s} \, .
\ee
On the other hand, from simple dimensional arguments, we
can find that the standard theory would predict
\be
\frac{|{\bf B}_g|}{\omega_{ring}} \Big|_{Th} 
\sim \frac{G_N m_{ring}}{c \, R} \sim 10^{-16} \, {\rm cm/s} \, ,
\ee
with $R = 7 \, {\rm cm}$ the radius of the ring.
The gravitomagnetic field of the niobium superconductive
ring should be some 18 orders of magnitude larger than
expected in order to account for the observed effect.

Lastly, anomalous large gravitomagnetic fields might be
produced by superfluid matter as well. Indeed, the authors 
of Ref.~\refcite{prop1} argue that such a possibility would be 
suggested by the unexplained generation and deletion of 
vortices in superfluids reported in~\refcite{hess}. So, roughly
speaking, coherent quantum systems would be able to produce 
anomalous strong gravitomagnetic fields. On the other
hand, other laboratory measurements confirm that the (gravitoelectric)
mass of superconductors does not change between the normal
and the superconductive state\cite{mass}: in other words,
coherent quantum systems seem to have standard gravitoelectric
properties but extraordinary gravitomagnetic ones. This 
consideration simplifies the picture and allows us for using 
the standard theory for the description of neutron star motion 
and for discussing separately the (unobserved) anomalous 
gravitomagnetic phenomena.

\section{Astrophysical Observations \label{s-astro}}

Neutron stars are the end-product of heavy stars after
supernova explosion\cite{glende}. Their mass is typically
about 1.5 Solar masses and their radius approximately 10 km: at
such high densities, matter would be made of neutrons, with 
a smaller fraction of protons and electrons. In the inner
part there could be also pions, kaons or free quarks.
Since we expect that the matter inside neutron stars is
at relatively low temperature, it is common belief that
neutrons form $^3P_2$ Cooper pairs and are in superfluid 
state and protons form $^1S_0$ Cooper pairs and are in
superconductive state\cite{sedrakian}. Like ordinary stars,
two neutron stars can form a binary system and, if at least 
one of them is detectable as radio pulsar, we can perform
high precision test of relativistic celestial mechanics. 
Indeed, typical orbital velocities are at least an order of 
magnitude larger than the ones of planets in the Solar System, 
they have very short orbital period, which is favorable for 
the observation of secular phenomena, and tidal effects can 
usually be ignored, since the orbital separation is much 
larger than their size. Today we know 5 neutron star--neutron
star binary systems which allow for high precision measurements
of this kind\cite{lattimer}.

The observable quantities of the binary systems are the
so-called ``non-orbital parameters'', 5 ``Keplerian parameters''
and 5 ``Post-Keplerian parameters'', all deduced by fitting
the arrival times of pulses. Since the Post-Keplerian parameters
depend on the theory of gravity, their measurement can be
used as testbed of general relativity\cite{damour}$^,$\cite{will}. In 
particular, in general relativity, for the case of negligible 
spin effects, the Post-Keplerian parameters depend only on 
the Keplerian ones and on the neutron star masses, implying 
that the measurement of 3 or more of them overconstraints 
the theory. At present, all the observations are in agreement
with the standard theory up to $v^2/c^2$ corrections and
effects of the gravitomagnetic fields are so far 
unobserved\cite{will} (but they may be in a near future\cite{p-p}). 
On the other hand, if matter in superfluid or superconductive state
generates gravitomagnetic field much larger than the one
produced by ordinary matter, 20 -- 30 orders of magnitude as 
possibly suggested by the unexplained laboratory experiments, the
mentioned self-consistent picture would be destroyed.

There are essentially three effects of a possible non-negligible
gravitomagnetic field in a binary system: the gravitomagnetic
precession of the pericenter, the gravitomagnetic time delay 
and the gravitomagnetic precession of the stellar spin. 
Unfortunately, all the observable parameters
of the binary system arise from data fitting, so the most
correct procedure would require to consider a particular 
theory (in our case a theory capable of producing anomalous
large gravitomagnetic fields) and find the new orbital
parameters. If the theory is overconstrained, observational
data may reject the model. On the other hand, the purpose
of the present work is to remain as general as possible,
showing that binary systems of compact stars disfavor anomalous
gravitomagnetic properties for superfluids and/or superconductors:
so, it suffices to require that the possible gravitomagnetic 
effect is not larger than the observed gravitoelectric one.
Moreover, laboratory experiments support the idea that only 
the gravitomagnetic fields of coherent quantum systems
are anomalous, whereas the gravitoelectric ones seem to be standard
at a high level of accuracy\cite{mass}, so we can reasonably
take the standard description of neutron star orbital motion
and discuss possible anomalous contributions, bearing in mind that
the theory of general relativity is in agreement with observations 
better than the percent\cite{will}.
Neglecting extraordinary fine tuning, this is a very
conservative picture and more particular cases certainly
are able to put much stronger constraints.

\subsection{Orbital precession}

The gravitoelectric orbital precession is
\be
\dot{\bar \omega}_{GE} = 3 \,
\left(\frac{T}{2 \pi}\right)^{-5/3}
\left(\frac{G_N M}{c^3}\right)^{2/3}
\left(1 - e^2\right)^{-1} \, ,
\ee
where $T$ is the orbital period and $M$ the total
mass of the system. For the known neutron star--neutron star binary
systems, we typically find 
$\dot{\bar \omega}_{GE} \sim 1-10 \, {\rm degrees/yr}$,
with a relative uncertainty in the range $10^{-3} - 10^{-6}$.
On the other hand, from Eq.~(\ref{o-precession}) one
find that the gravitomagnetic contribution in general
relativity is
\be
\dot{\bar \omega}_{GM} \sim 
\frac{G_N  J}{c^2 \, a^3} 
\sim 10^{-5} \, {\rm degrees/yr} \, .
\ee
Here I took $J \sim M_\odot \, R^2_{NS} \, \omega_{NS}$, with 
$R_{NS} \sim 10 \, {\rm km}$ the neutron star radius and
$\omega_{NS} \sim 100 \, {\rm rad/s}$ the neutron star
rotation frequency, and $a \sim 10^{11} \, {\rm cm}$
as standard stellar separation distance.
If neutrons in superfluid state and proton in the 
superconductive one produced a gravitomagnetic field just $\sim 5$ 
orders of magnitude stronger than the expected one, the 
gravitomagnetic orbital precession contribution
would be at the same level of the gravitoelectric effect, 
with disagreement between theory and observational data. Of course
one can reasonably argue that the fraction of the star in 
superconductive and superfluid state is model dependent
and that the fraction of coherent quantum matter
may be as low as some percent. However, even if this
is certainly possible, such a consideration can
only change of 2 or 3 orders of magnitude our previous
estimates and can hardly save the idea of strong gravitomagnetic
fields from superfluids and superconductors.

\subsection{Time delay}

Let us now turn to the gravitational time delay, that is
the time delay of light rays emitted by the pulsar when they pass
near the other neutron star, with respect to the case the 
spacetime was flat. The gravitoelectric effect is the 
well-known Shapiro time delay. In the case of a binary
system, we can quantify the phenomenon considering the 
gravitational delay of the light when the pulsar is in
front of and behind the companion 
\be\label{shapiro-1}
\Delta t_{GE} = \frac{2 G_N m}{c^3}
\, \ln \left(\frac{4 a^2}{d^2}\right) \, ,
\ee
where $m$ is the mass of the companion and $d$ the impact 
parameter of the light ray, which we can take of order of the 
radius of the neutron star companion. On the other hand,
the gravitomagnetic effect is given by\cite{gem-delay}
\be\label{shapiro-2}
\Delta t_{GM} = - \frac{2 G_N}{c^4} 
\frac{{\bf J} \cdot \hat{\bf n}}{d} \, ,
\ee
where $\hat{\bf n}$ is the unit vector normal to the plane
formed by ${\bf x}_1$ and ${\bf x}_2$ and directed along 
${\bf x}_1 \wedge {\bf x}_2$, ${\bf x}_1$ and ${\bf x}_2$ being the 
position vectors respectively of the pulsar and of the observer
with respect to the rotating body.
Plugging typical binary system parameters into Eqs.~(\ref{shapiro-1})
and (\ref{shapiro-2}), we find a gravitoelectric time delay
at the level of 0.1~ms and a gravitomagnetic contribution of
about $10^{-5}$~ms. If the superfluid neutrons and
the superconductive protons produced a gravitomagnetic
field $\sim 5$ orders of magnitude stronger than ordinary matter, 
once again the successful picture would be destroyed. Moreover, 
the effect could be somehow well recognizable, because while 
the gravitoelectric contribution depends only on the impact
parameter $d$, the gravitomagnetic one becomes positive or
negative, depending on the sign of ${\bf J} \cdot \hat{\bf n}$.

\subsection{Spin precession}

The last effect we can consider is the spin precession.
The geodesic (or de Sitter) precession depends on the velocity
of the spinning test-particle ${\bf v}$ and on the gradient
of the gravitational potential $\Phi_g$, which is related
to the gravitoelectric field ${\bf E}_g$ by
\be
{\bf E}_g = - \nabla \Phi_g 
- \frac{1}{c} \frac{\partial}{\partial t} 
\left(\frac{1}{2} {\bf A}_g\right) \, .
\ee
On the other hand, it is independent of the gravitomagnetic
field of the source ${\bf B}_g$. The geodesic angular frequency 
is\cite{will}
\be
{\bf \Omega}_{G} = \frac{3}{2} {\bf v} \wedge \nabla \Phi_g \, .
\ee
A rough estimate is
\be
\Omega_G \sim \frac{G_N \, M \, v}{c^2 \, r^2} 
\sim 10^{-2} \, {\rm rad/yr} \, ,
\ee
where $v \sim 300 \, {\rm km/s}$ is the typical neutron star
velocity. Such a simple evaluation is compatible with 
observations\cite{spin}. On the other hand, the contribution
of the gravitomagnetic field of the neutron star companion
in the standard theory is the Lense-Thirring precession,
which can be evaluated
\be
\Omega_{LT} \sim \frac{G_N}{c^2} \frac{J}{r^3}
\sim 10^{-7} \, {\rm rad/yr} \, .
\ee
Even in this third case, the gravitomagnetic effect is expected
to be about 5 orders of magnitude smaller than the gravitoelectric
one, so that it is quite improbable that superfluids and superconductors
produce a gravitomagnetic field 20 -- 30 orders of magnitude larger than
the one of ordinary classical matter.

\section{Conclusions \label{s-conclusion}}

The possibility of anomalous strong gravitomagnetic
fields from coherent quantum system has been considered
by various authors. In particular there are three
unexplained observed phenomena occurring in experiments
involving rotating superconductive rings and one involving
superfluid matter that may support this fascinating idea. 
On the other hand, general relativity is well tested only 
for classical macroscopic bodies like planets or satellites, 
so that new gravitational properties of non-common matter 
is certainly worthy of investigation.

However, in the commonly accepted picture, neutron stars
are made of neutrons in superfluid state and proton in
superconductive one, with also the possibility of 
superfluid hyperons and quarks. Since we know 5 neutron 
star--neutron star binary systems where at least one of the 
star is a radio pulsar and where we can perform high precision
tests of celestial mechanics, in principle we would be able to check
the hypothesis of the anomalous strong gravitomagnetic field
produced by superfluids and superconductors.
The standard theory works perfectly, while the possible
anomalous large gravitomagnetic field suggested by laboratory
experiments would make gravitomagnetic effects larger than
the gravitoelectric ones. Even if the amount of matter
in superfluid and superconductive state is model dependent,
the laboratory experiments require so strong gravitomagnetic
fields that we can at least conclude that present knowledge 
of neutron star physics strongly disfavors the idea
suggested in Refs.~\refcite{prop1} and \refcite{prop2} as solution to
the unexplained results in Refs.~\refcite{london} and \refcite{tajmar1}.

\section*{Acknowledgments}
I wish to thank Alessandro Drago for useful discussions on
the physics of compact stars.
This work was supported in part by NSF under grant PHY-0547794 
and by DOE under contract DE-FG02-96ER41005.

\end{document}